# Optical phonons in new ordered perovskite $Sr_2Cu(Re_{0.69}Ca_{0.31})O_y$ system observed by infrared reflectance spectroscopy

J. Ahmad[*, a], M. Shizuya[b], M. Isobe[b], S. Alam[b] and H. Uwe[a]

[a] Institute of Materials Science, University of Tsukuba, Tsukuba, Ibaraki, 305-8573, Japan.
[b] National Institute for Materials Science, 1-1 Namiki, Tsukuba, Ibaraki 305-0044, Japan.

## Abstract

We report infrared reflectivity spectra for a new correlated cupric oxide system $Sr_2Cu(Re_{0.69}Ca_{0.31})O_y$ with $y \sim 0.6$ at several temperatures ranging between 8 and 380 K. The reflectivity spectrum at 300 K comprises of several optical phonons. A couple of residual bands located around 315 and 653 cm$^{-1}$ exhibit exceptionally large intensity as compared to the other ones. The overall reflectivity spectrum lifts up slightly with increasing temperature. The energy and damping factor of transverse-optical phonons are determined by fitting the imaginary dielectric constant $\varepsilon_2$ by Lorentz oscillator model and discussed as a function of temperature in terms of lattice anharmonicity.

*Corresponding author.

Dr. Javed Ahmad

Postal Address: Institute of Materials Science, University of Tsukuba, Tsukuba, Ibaraki, 305-8573, Japan

Phone: +81-29-853-5141

Fax: +81-29-853-6001

Email: dr.j.ahmad@gmail.com

## 1. Introduction

The ordered perovskite $Sr_2Cu(Re_{0.69}Ca_{0.31})O_6$ (SCRCO) is a new cupric oxide system, which exhibits a strong magnetic correlation in its ferromagnetic (ferrimagnetic) phase. The ferromagnetic behavior is observed below $T_C$ ~ 440 K with spontaneous magnetization of about 1 $\mu_B$ per formula unit at 5 K [1]. The proposed ordered perovskite-type structure with a cubic symmetric unit lattice, space group $Pm\overline{3}m\,(O^1{}_h)$, consists of eight perovskite-like subcell blocks. In each subcell, the Sr ions exclusively occupy the A-site and the three small cations Cu, Re, and Ca are selectively located at the B-sites in an ordered fashion. There are two different kinds of Cu sites: Cu1 and Cu2 with valences 3+ and 2+, respectively [2,3]. The Cu1, Cu2 and $O_2$ ions form a 3-D frame network connected by Cu-O-Cu straight bond. On the basis of the results of electron energy loss spectroscopy, Isobe *et al*. proposed the magnetic origin of this system to be due to the Cu spin moment and not the Re ones, which are non-magnetic, a state with valency 7+ [2]. They thought that the high $T_C$ probably originates in the strong antiferromagnetic coupling between Cu spins via an O ion in the $180^o$ Cu-O-Cu bonding network, a situation similar to that in cuprate superconductors.

In this paper, we report the preliminary spectra of infrared reflectivity recorded at various temperatures ranging between 8 and 380 K for ordered perovskite SCRCO in order to obtain information on the charge carrier dynamics and to study the infrared-active phonons. To our knowledge, there is no such study reported so far for SCRCO system.

## 2. Experiment

The ordered perovskite SCRCO can only be prepared under high temperature/high pressure conditions. A polycrystalline sample was prepared by solid-state reaction using high-pressure synthesis technique from mixtures of $SrO_2$, $Ca_2CuO_3$, CuO, $ReO_3$, and metal Re powder by a research group at NIMS, Japan [1,4]. The powder X-ray diffraction ensured almost pure single-phase sample within an impurity level less than ~ 1%. The grain morphology was measured using a JEOL JXA-8621 scanning electron microscope. The average grain size was 1-5 $\mu$m.

For the reflectivity measurements, one of the surfaces of the circular disk shaped sample was mechanically polished to almost a mirror like one, first with 0.3$\mu$m and then with 0.05 $\mu$m $Al_2O_3$ powder. The reflectivity spectra were measured in nearly normal incidence configuration using a Bruker 66v FTIR spectrometer in the frequency range 30-7500 $cm^{-1}$ for temperature interval 8-380 K. For higher frequency range 5000-30000 $cm^{-1}$ a grating type spectrometer was used and reflectivity spectrum was recorded only at room temperature. Since the background reflectivity did not show significant temperature dependence, we attached the high frequency spectrum at room temperature with the low frequency spectra for all measured temperatures in order to apply Kramers Kronig (KK) transformation. In this way, we could determine accurately the optical dielectric constant, $\varepsilon$ ($\omega$), by applying KK transformation to the reflectivity spectra for such a broad frequency range. Further detail of the experimental set up and reflectivity measurements can be seen elsewhere [5-7].

### 3. Results and Discussion

The reflectivity spectra for SCRCO in the frequency interval 30-1000 $cm^{-1}$ are shown in Fig. 1. The spectra consist of several residual bands with two having appreciably large intensity. Below ~ 100 $cm^{-1}$, several kinks appeared that could not be resolved from the increased noise and it is, therefore, not clear if these are the extra residual bands at low frequency. The anomalous increase of reflectivity spectra below ~ 100 $cm^{-1}$ seems not to be of a Drude type as the increasing trend continues and can be seen even at the lowest temprature, i.e., 8 K. However, the rest of the spectra is of a typical insulator or semiconductor. As for temperature dependence, we find a little broadening of the bands and lifting up slightly of the overall reflectivity spectra with increasing temperature up to 380 K. The slight lifting up of spectra may indicate the presence of small number of free carriers, which are thermally excited with increasing temperature and contribute to the background reflectivity. In polycrystalline samples of the grain size with 1-5 $\mu$m, the Mie scattering from the grains hampers observation of the bulk spectrum at the wavelength around the grain size. This means the spectra at above ~ 2000 $cm^{-1}$ may not provide reliable information about the system. However, at frequency well below 2000 $cm^{-1}$, the problem cannot arise as the wavelength of incident electromagnetic radiation is large enough to spread over several grains. We, therefore, believe that the observed phonon spectra reflect a bulk property and its temperature dependence is an intrinsic behavior in the frequency range 30-1000 $cm^{-1}$.

Figure 1 (lower panel) shows the imaginary part of the dielectric constant ($\varepsilon_2$) of the corresponding reflectivity spectra obtained by the KK transformation. It can be seen that the phonon modes are well resolved even in this polycrystalline sample, which seems to be resulted from the cubic symmetry of the crystal lattice. Experimental observation of large number of transverse-optical (TO) phonons is consistent with the complicated structure of the system where the unit cell consists of four formula units. Here, the peak positions measure the resonant frequency of the TO phonons. For the quantitative analysis of the TO phonons, we have fitted the $\varepsilon_2$ spectra with Lorentz oscillator model in which the dielectric function $\varepsilon(\omega)$ is defined as;

$$\varepsilon(\omega) = \varepsilon_\infty + \sum_j \frac{\omega_{\mathrm{TO}}^2(j) S_j}{\omega_{\mathrm{TO}}^2(j) - \omega^2 - i\omega\gamma_j}, \qquad (1)$$

where $\varepsilon_\infty$ is the high-frequency dielectric constant, $\omega_{\mathrm{TO}}(j)$ the TO phonon frequency, $S_j$ the oscillator strength, $\gamma_j$ the phonon-damping factor, and *j* represents the total number of TO phonons appeared in the spectrum. We have assumed 12 phonons, i.e., $j$ = 1-12, in the frequency range 80-1000 cm$^{-1}$ in order to reproduce the experimentally obtained $\varepsilon_2$ spectra. It can be seen in Fig. 1 and 2 that the high-frequency phonon modes are well resolved even in this polycrystalline sample. We focus on how these well resolved phonon modes are affected by varying temperature.

In Fig. 2, we plot $\omega_{\mathrm{TO}}(j)$ extracted from the Lorentzian fit to the phonon spectra as a function of temperature. All modes become soft except mode no. 11, which hardens with increasing temperature. Mode hardening can be understood in terms of lattice anhamonicity. Thermal expansion is the most familiar phenomenon associated with the anharmonic terms. As temperature rises, the amplitude of lattice vibration increases so that the average root mean square value of the displacement from the equilibrium position increases. The anharmonic terms contribute to the free energy of the crystal. The whole crystal then expands until it finds the volume where total free energy is minimum.

The frequency of lattice modes is a function of volume and thus change in volume gives rise to the same relative change in the frequency of every mode [8]. One of the possible explanations for the increase in frequency of the mode 11 with temperature is the phonon cloud contribution to the frequency as in the case of the ferroelectric soft phonon.

In Fig. 3, the damping factor $\gamma_j$ is plotted as a function of temperature. As usual, the $\gamma_j$ increases with temperature for all modes except mode no. 9, the $\gamma_j$ of which anomalously decreases. Mode 9 is a weak mode with very less intensity even at the lowest temperature 8 K and lies in between two relatively large-intensity modes 8 and 10. On increasing temperature, the widths of the modes 8 and 10 increase significantly and it becomes difficult experimentally to accurately distinguish mode 9 with its intensity and width decreasing due to increased error in the fit parameters. The broadening of the line width with temperature may also be the reason of the lifting up of the reflectivity as well as $\varepsilon_2$ spectra at high temperature. Moreover, A peak (not shown here) corresponding to electronic excitations has been observed around 2580 $cm^{-1}$ in the $\varepsilon_2$ spectrum of 8K, which shifts toward low frequency and seems to transfer its spectral weight to lower frequency region on increasing temperature. This electronic contribution to background reflectivity of phonons may also result in lifting up the spectra.

## 4. Conclusion

We have reported the experimental results of the optical reflectivity of the new ordered perovskite $Sr_2Cu(Re_{0.69}Ca_{0.31})O_6$ polycrystal. The reflectivity spectra are found to be of a typical insulator comprising of several residual bands. The frequency of the several experimentally observed transverse optical phonons is determined by using the Lorentz oscillator model. The TO phonon frequency of most of the phonons is found to decrease with increasing temperature, though the mode at 586.5 $cm^{-1}$ at 8 K exhibits hardening with temperature.

**Acknowledgements**

We thank N. Nishida, the Chemical Analysis Center, for help in electron microscopy for determining the grain size. J. Ahmad would like to thank the postdoctoral fellowship program of the University of Tsukuba for financial support.

**References**

[1] E. Takayama-Muromachi, T. Drezen, M. Isobe, N. D. Zhigadlo, K. Kimoto, Y. Mitsui, E. Kita, J. Solid State Chem. 170 (2003) 24.

[2] M. Isobe, Y. Uchida, K. Kimoto, M. Arai, E. Takayama-Muromachi, J. Magn. Magn. Mater. 272-274 (2004) 623.

[3] Y. Uchida, M. Isobe, E. Takayama-Muromachi, J. Magn. Magn. Mater. 272-274 (2004) 818.

[4] M. Isobe, T. Kawashima,K. Kosuda, Y. Matsui, E. Takayama-Muromachi, Physica C 234 (1994) 120.

[5] J. Ahmad, T. Nishio, H. Uwe, Physica C 388-389 (2003) 455.

[6] J. Ahmad, H. Uwe, Physica C 412-414 (2004) 288.

[7] J. Ahmad, H. Uwe, Phys. Rev. B 72 (2005) 125103.

[8] J. M. Ziman, in: $2^{nd}$(Eds.), Principles of the theory of Solids, Cambridge University Press, 1972, pp. 66.

**List of Figures**

Fig. 1. The optical reflectivity spectra (upper) and the corresponding imaginary part of dielectric constant $\varepsilon_2$ of $Sr_2Cu(Re_{0.69}Ca_{0.31})O_6$ along with the best fit of Lorentz Oscillators (lower) are shown at two end temperatures. The peak positions give the energy of TO phonons. Note the number assigned to each phonon mode.

Fig. 2. The TO-phonon energy, $\omega_{\mathrm{TO}}(j)$ as a function of temperature extracted from the Lorentzian fit to the $\varepsilon_2$. The $\omega_{\mathrm{TO}}(j)$ of only the well resolved phonon modes is plotted. The increase in $\omega_{\mathrm{TO}}(9)$ at 380 K is due to error in fit as the mode is hardly resolved at high temperature.

Fig. 3. The TO-Phonon damping factor, $\gamma_j$ as a function of temperature extracted from the Lorentzian fit to the $\varepsilon_2$. The $\gamma_j$ of only the well-resolved phonon modes is plotted. Error bars show error in the fit parameters.

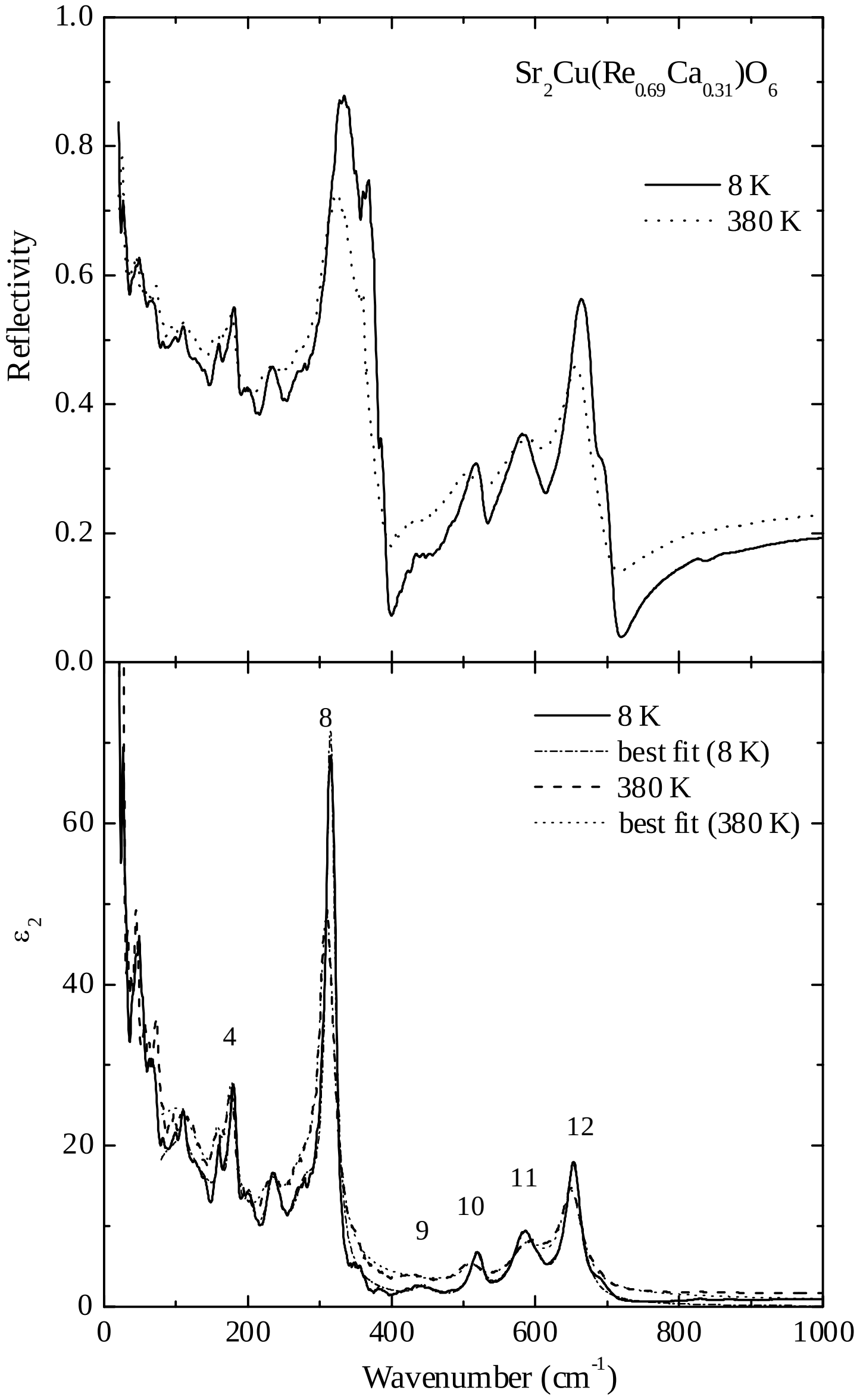

Fig. 1

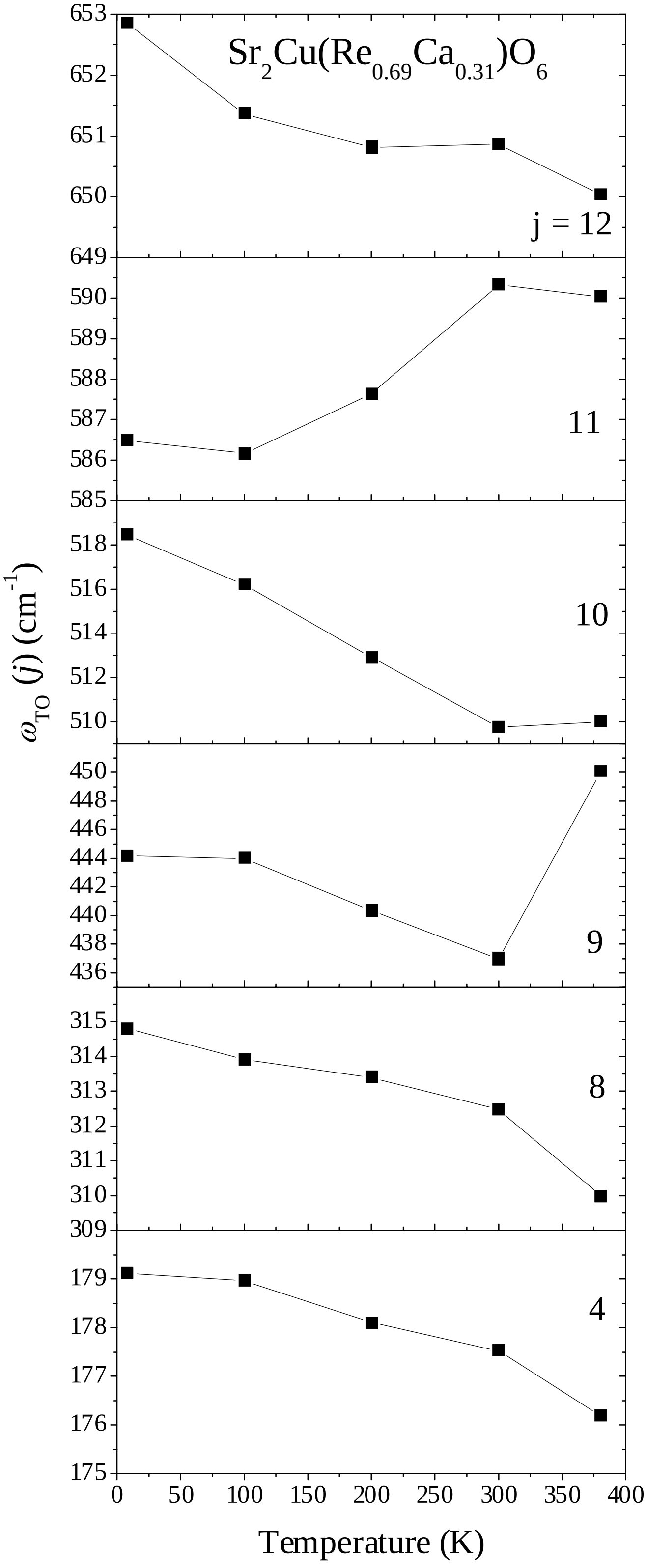

Fig. 2

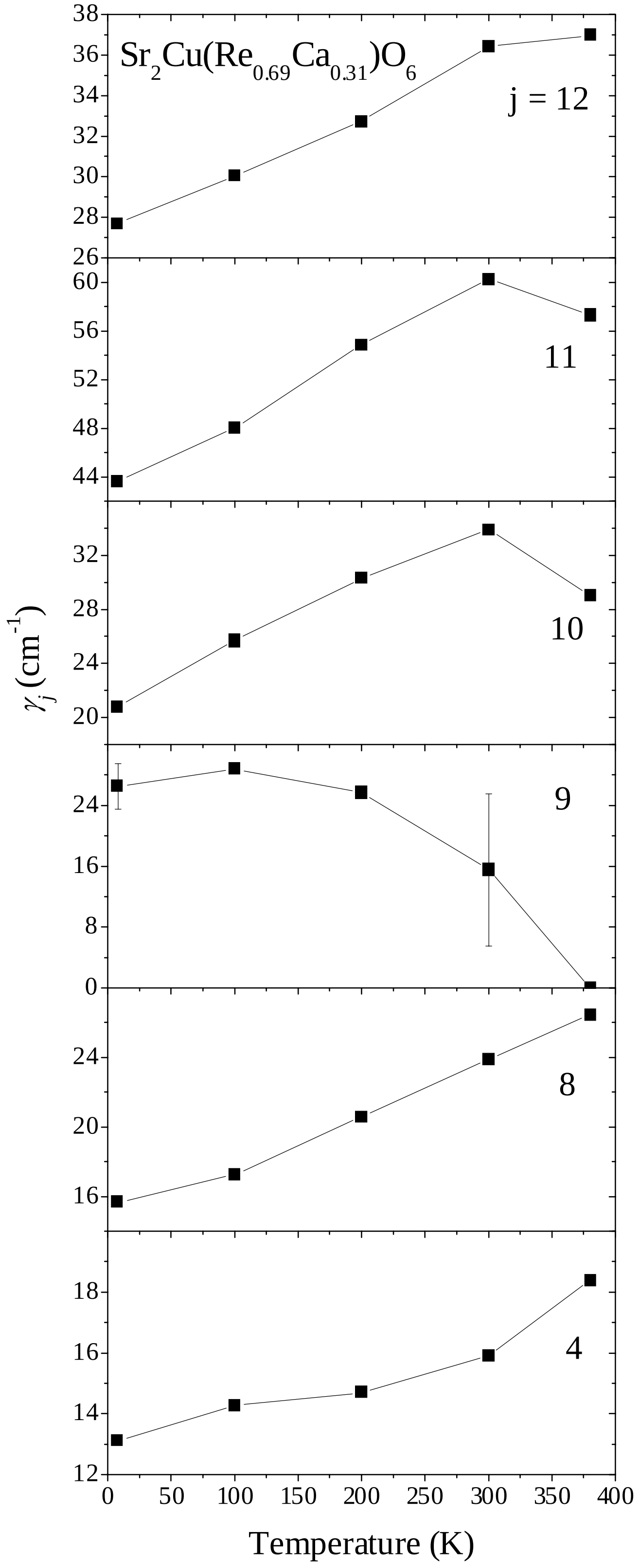

Fig. 3